\documentclass[aps,prb,twocolumn,amsmath,amssymb,nofootinbib,
  superscriptaddress]{revtex4-2}

\usepackage{graphicx}
\usepackage{bm}
\usepackage{amsmath}
\usepackage{amssymb}
\usepackage{physics}
\usepackage[colorlinks=true,linkcolor=blue,urlcolor=blue,citecolor=blue]{hyperref}
\usepackage[usenames,svgnames]{xcolor}
\usepackage{makecell}
\usepackage[normalem]{ulem}

\newcommand{\gcc}[1]{\textcolor{black}{#1}}

\newcommand{\Hop}{\hat{H}}
\newcommand{\Uop}{\hat{U}}
\newcommand{\Top}{\mathcal{T}}
\newcommand{\Himp}{\hat{H}_{\rm imp}}
\newcommand{\gHimp}{\mathcal{H}_{\rm imp}}
\newcommand{\Hint}{\hat{H}_{\rm int}}
\newcommand{\Hbath}{\hat{H}_{\rm bath}}
\newcommand{\Hhyb}{\hat{H}_{\rm hyb}}

\newcommand{\aop}{\hat{a}}
\newcommand{\adop}{\hat{a}^{\dagger}}

\newcommand{\cop}{\hat{c}}
\newcommand{\cdop}{\hat{c}^{\dagger}}
\newcommand{\hc}{{\rm H.c.}}
\newcommand{\rhoop}{\hat{\rho}}
\newcommand{\Zimp}{Z_{{\rm imp}}}

\newcommand{\gK}{\mathcal{K}}
\newcommand{\gI}{\mathcal{I}}
\newcommand{\bolda}{\bm{a}}
\newcommand{\boldabar}{\bar{\bm{a}}}
\newcommand{\abar}{\bar{a}}
\newcommand{\im}{{\rm i}}
\newcommand{\contour}{\mathcal{C}}
\newcommand{\gA}{\mathcal{A}}

\newcommand{\branch}{\mathcal{C}}

\newcommand{\EqDef}{\stackrel{\mathrm{def}}{=}}

\newcommand{\hnu}{Key Laboratory of Low-Dimensional Quantum Structures and Quantum Control of Ministry of Education, Department of Physics and Synergetic Innovation Center for Quantum Effects and Applications, Hunan Normal University, Changsha 410081, China
}

\newcommand{\ustc}{Key Laboratory of Precision and Intelligent Chemistry, University of Science and Technology of China, Hefei 230026, China}

\begin{document}

\title{Infinite Grassmann time-evolving matrix product operators for quantum impurity problems after a quench}

\author{Zhijie Sun}
\affiliation{\ustc}

\author{Ruofan Chen}
\affiliation{College of Physics and Electronic Engineering, and Center for Computational Sciences, Sichuan Normal University, Chengdu 610068, China}

\author{Zhenyu Li}
\email{zyli@ustc.edu.cn}
\affiliation{\ustc}

\author{Chu Guo}
\email{guochu604b@gmail.com}

\affiliation{\hnu}



\begin{abstract}
An emergent numerical approach to solve quantum impurity problems is to encode the impurity path integral as a matrix product state. For time-dependent problems, the cost of this approach generally scales with the evolution time. Here we consider a common non-equilibrium scenario where an impurity, initially in equilibrium with a thermal bath, is driven out of equilibrium by a sudden quench of the impurity Hamiltonian. Despite that there is no time-translational invariance in the problem, we show that we could still make full use of the infinite matrix product state technique, resulting in a method whose cost is essentially independent of the evolution time.
We demonstrate the effectiveness of this method in the integrable case against exact diagonalization, and against existing calculations on the L-shaped Kadanoff-Baym contour in the general case. Our method could be a very competitive method for studying long-time non-equilibrium quantum dynamics, and be potentially used as an efficient impurity solver in the non-equilibrium dynamical mean field theory.
\end{abstract}

\maketitle


\section{Introduction}

Strongly correlated electron systems driven out of equilibrium underlie a variety of exotic phenomena that are not easily related to equilibrium physics~\cite{OriSig2018, DieJan2018, LigAvi2018, BluOmr2021, SkoSko2021, MakWin2021, BaoTan2022}. 
However, accurate numerical solutions to these problems remains challenging. 
A particular feature of the non-equilibrium scenario is the slow relaxation time that could often be orders of magnitude larger than the characteristic time scale of the system, which makes it even harder to solve compared to the equilibrium counterpart~\cite{AokiWerner2014, EisFri2015, Mar2018, Witra2018}.

The Anderson impurity model (AIM) is an prototypical model for studying strongly correlated effects, which describes a localized electron immersed in a bath of itinerant electrons~\cite{anderson1961-localized}. Despite the simplicity of the model, solving the long-time dynamics of it still poses great numerical challenge. Existing methods include exact diagonalization (ED)~\cite{CaffarelKrauth1994,KochGunnarsson2008,GranathStrand2012,LuHaverkort2014,ZaeraLin2020,HeLu2014,HeLu2015}, time-evolving matrix product state (MPS)~\cite{WolfSchollwock2014b,GanahlEvertz2014,GanahlVerstraete2015,WolfSchollwock2015,GarciaRozenberg2004,NishimotoJeckelmann2006,WeichselbaumDelft2009,BauernfeindEvertz2017,LuHaverkort2019,WernerArrigoni2023,KohnSantoro2021,KohnSantoro2022}, numerical renormalization group~\cite{Wilson1975,Bulla1999,BullaPruschke2008,Frithjof2008,ZitkoPruschke2009,DengGeorges2013,StadlerWeichselbaum2015,LeeWeichselbaum2016,LeeWeichselbaum2017}, the hierarchical equation of motion~\cite{YoshitakaKubo1989,jin2007-dynamics,jin2008-exact,yan2016-dissipation,cao2023-recent}, and recent advanced quantum Monte Carlo methods~\cite{CohenGull2014,CohenMillis2014,CohenMillis2015,ChenReichman2017a,ChenReichman2017b,ErpenbeckCohen2023}. These method could give satisfactory results in various model settings, but none can provide generally stable and accurate solutions.

An emergent numerical approach to solve the AIM on the real-time axis is to encode the impurity path integral (PI) as an MPS, which is fundamentally different from conventional MPS methods where MPS is used to represent the impurity-bath wave function at a particular time. Two variants in this category have been proposed up to date. One uses fermionic MPS in the Fock state basis~\cite{ThoennissAbanin2023b,NgReichman2023,NayakWerner2025}. The other uses Grassmann MPS (GMPS) in the coherent state basis, which will be referred to as the Grassmann time-evolving matrix product operator method (GTEMPO)~\cite{ChenGuo2024a,XuChen2024} due to its close relation to the time-evolving matrix product operator method for bosonic impurity problems~\cite{StrathearnLovett2018}. 
The major advantage of the PI-based MPS methods, compared to the conventional wave-function based MPS methods, is that the bath degrees of freedom are integrated out analytically via the Feynman-Vernon influence functional (IF)~\cite{FeynmanVernon1963}. Therefore the PI-based MPS methods are free of bath discretization error and could potentially be more efficient.

In this work we aim to extend the GTEMPO method to solve the non-equilibrium real-time dynamics of the AIM. 
We will further focus on a common non-equilibrium scenario where only the impurity Hamiltonian is time-dependent (e.g., the localized electron is under some external driving).
In the PI formalism, the non-equilibrium setup is usually formulated on the L-shaped Kadanoff-Baym contour~\cite{kadanoff1962-quantum}, which has been studied using the GTEMPO method~\cite{ChenGuo2024g}. However, as finite GMPSs are used in GTEMPO, its cost scales with both the real and imaginary times.

Here we explore an equivalent formulation of the impurity problem on the real-time axis only, which restores the time-translational invariance of the Feynman-Vernon IF. In combination with a specially designed quench protocol for the time-dependent impurity Hamiltonian, we could make full use of the well-established infinite MPS technique, resulting in a non-equilibrium infinite GTEMPO (neq-iGTEMPO) method whose computational cost is essentially independent of the evolution time.
We demonstrate the effectiveness of this method in the integrable case against exact diagonalization, and against existing GTEMPO calculations on the L-shaped Kadanoff-Baym contour in the interacting case. These results show that our method could be very competitive for studying long-time non-equilibrium quantum dynamics of the AIM, and be potentially used as an efficient impurity solver in the non-equilibrium dynamical mean field theory (DMFT).

\section{General description of non-equilibrium quantum dynamics}\label{sec:PI}

\begin{figure}
  \includegraphics[width=\columnwidth]{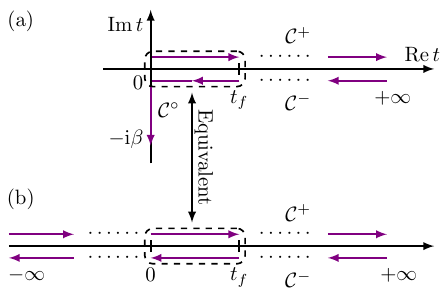} 
  \caption{Two equivalent descriptions of non-equilibrium real-time quantum dynamics under a time-dependent Hamiltonian $\Hop(t)$ and starts from the equilibrium state in Eq.(\ref{eq:thermalstate}) at $t=0$. (a) The evolution is naturally formulated on the L-shaped Kadanoff-Baym contour, which consists of the forward ($\contour^+$), backward ($\contour^-$) and imaginary ($\contour^o$) branches, corresponding to Eq.(\ref{eq:dynamics1}). (b) An equivalent formulation of the problem on the Keldysh contour with the time-dependent Hamiltonian $\Hop^{\eta}(t)$ in Eq.(\ref{eq:Heta}) that starts from $t = -\infty$, which consists of the forward and backward branches only, corresponding to Eq.(\ref{eq:dynamics2}). The dashed box in both panels indicates the time window within which we would like to calculate any observables.
    }
    \label{fig:fig1}
\end{figure}

We first briefly review the general description of non-equilibrium quantum dynamics under a time-dependent Hamiltonian $\Hop(t)$, which is used in the path integral formalism.
The initial state of the whole system is assumed to be in the thermal state with inverse temperature $\beta$, denoted as
\begin{align}\label{eq:thermalstate}
\rhoop^{\rm th} = e^{-\beta \Hop(0)}.
\end{align}
The real-time dynamics is described by 
\begin{align}\label{eq:dynamics1}
\rhoop(t) = \Uop(t, 0) \rhoop^{\rm th} \Uop(0, t),
\end{align}
where $\Uop(t, 0) =\Top \{e^{-\im \int_0^t d\tau \Hop(\tau)}\} $ is the evolutionary operator with $\Top$ the time-ordered operator, and $\Uop(0, t) = \Uop^{\dagger}(t, 0)$. The partition function, which is the central quantity to evaluate for computing any observables, can be calculated as 
\begin{align}\label{eq:partition}
Z(t) = \tr\left[\rhoop(t)\right] = \tr\left[ e^{-\beta \Hop(0)} \Uop(0, t) \Uop(t, 0)\right],
\end{align}
where we have used the cyclic property of trace in the second equality. Reading the expression inside the square bracket in Eq.(\ref{eq:partition}) from right to left, the evolution can be naturally understood as along an L-shaped contour: it starts from time $0$ to $t$ by a forward evolution $\Uop(t, 0)$, then it turns back from $t$ to $0$ by a backward evolution $\Uop(0, t)$, finally it ends with an imaginary time evolution from $0$ to $-\im \beta$. 
This contour is usually referred to as the Kadanoff-Baym contour~\cite{kadanoff1962-quantum}, which is depicted in Fig.~\ref{fig:fig1}(a). 
The three stages of evolution are referred to as the \textit{forward} ($\contour^+$), \textit{backward} ($\contour^-$) and \textit{imaginary} ($\contour^o$) branches respectively.

The mixture of real- and imaginary-time evolution is not convenient for numerical calculations. 
A majority of numerical studies for non-equilibrium quantum dynamics are thus restricted to the real-time axis only.
The origin of the inconvenience is that the initial state $\rhoop^{\rm th}$ is generally an entangled state which does not allow an easy preparation. Nevertheless, there is another equivalent formulation which calculates $Z(t)$ based on the forward and backward branches only (e.g., the Keldysh contour). The idea is sketched in the following. 

We first rewrite the time-dependent Hamiltonian as
\begin{align}
\Hop(t) = \Hop_0  + \Hop_1(t),
\end{align}
where $\Hop_0$ is some ``simple'' Hamiltonian for which the corresponding thermal state can be easily prepared (e.g., some non-interacting Hamiltonian), and $\Hop_1(t) = \Hop(t) - \Hop_0$. 
Now, to prepare $\rhoop^{\rm th}$ at time $t=0$, we can start from $t = -\infty$ with another initial state $\rhoop^{\rm th}_{-\infty} = e^{-\beta \Hop_0}$ and adiabatically switch on $\Hop_1(t)$ with the protocol
\begin{align}\label{eq:Heta}
\Hop^{\eta}(t) =\Hop_0  + \eta(t) \Hop_1(t),
\end{align}
where $\eta(t)$ is a slowly varying function from $0$ to $1$, with $\eta(-\infty)=0$ and $\eta(t)=1$ for $t\geq 0$, then at time $t=0$, we have
\begin{align}\label{eq:adiabatic}
\rhoop^{\eta}(0) = \Uop^{\eta}(0, -\infty) \rhoop^{\rm th}_{-\infty} \Uop^{\eta}(-\infty, 0) \propto \rhoop^{\rm th}.
\end{align}
The last relation in Eq.(\ref{eq:adiabatic}) holds provided that the \textit{adiabatic assumption is not violated}, e.g., there is no level degeneracy and level crossing~\cite{StefanucciLeeuwen2013}. The overall dynamics in Eq.(\ref{eq:dynamics1}) can thus be equivalently written as
\begin{align}\label{eq:dynamics2}
\rhoop(t) \propto \rhoop^{\eta}(t) = \Uop^{\eta}(t, -\infty) \rhoop^{\rm th}_{-\infty} \Uop^{\eta}(-\infty, t),
\end{align}
which is depicted in Fig.~\ref{fig:fig1}(b).
We have also extended the final time to $+\infty$ in both panels of Fig.~\ref{fig:fig1} by making use of the causality property of real-time evolution (e.g., the future dynamics does not affect the past observables). Here we note that the adiabatic assumption may not hold in general, and the equivalence between Eq.(\ref{eq:dynamics1}) and Eq.(\ref{eq:dynamics2}) needs to be checked case by case.

\section{The non-equilibrium quantum impurity problem}
Now we specialize to the quantum impurity problems (QIPs) driven out of equilibrium. The total Hamiltonian can be generally written as
\begin{align}
\Hop(t) = \Himp(t) + \Hhyb(t) + \Hbath,
\end{align}
with $\Himp(t)$ the impurity Hamiltonian, $\Hhyb(t)$ the hybridization Hamiltonian which describes the coupling between the impurity and the bath, $\Hbath$ the bath Hamiltonian which is usually assumed to be non-interacting and time-independent. 

Two setups are often considered for the non-equilibrium dynamics of QIPs: (1) The initial state is a separable state of the impurity and the bath, e.g., $\rhoop(0) = \rhoop_{\rm imp} \otimes \rhoop_{\rm bath}^{\rm th}$ where $\rhoop_{\rm imp}$ is some arbitrary impurity state and $\rhoop_{\rm bath}^{\rm th} = e^{-\beta \Hbath}$;
(2) The initial state is the thermal state in Eq.(\ref{eq:thermalstate}). The first setup is often used to model open quantum dynamics and non-Markovian effects, where the impurity gradually builds up entanglement with the bath. Since in this setup the initial state is separable (thus can be easily prepared), it can be directly formulated on the keldysh contour and thus solved using both conventional wave-functional based MPS methods~\cite{KohnSantoro2022} and the PI-based MPS methods~\cite{ThoennissAbanin2023b,ChenGuo2024a,ChenGuo2024c}. 
The second setup is often used as an intermediate impurity problem to be solved in the non-equilibrium DMFT~\cite{AokiWerner2014}. As the initial state is entangled, it is usually formulated on the Kadanoff-Baym contour in the PI formalism. The GTEMPO method has been used to solve the non-equilibrium AIM in the second setup on the Kadanoff-Baym contour, where finite GMPSs are used to encode the whole information on this contour~\cite{ChenGuo2024g}.
Therefore its computational cost to build the GMPSs scales at least linearly against both the real time $t$ and the imaginary time $\beta$ (the scaling is linear if the non-Markovian memory size does not scale with time~\cite{Guo2022c}). 

In this work we focus on the second setup. We will further assume $\Hhyb$ to be time-independent and denote $\Hint = \Hhyb + \Hbath$ which contains all the bath effects. Instead of solving the problem on the L-shaped Kadanoff-Baym contour in Fig.~\ref{fig:fig1}(a), we will extend the GTEMPO method to solve the problem on the equivalent Keldysh contour in Fig.~\ref{fig:fig1}(b). A direct advantage on the Keldysh contour is that the Feynman-Vernon IF is time-translationally invariant, as $\Hint$ is time-independent, therefore it can be represented as an infinite GMPS. Drawing connection to Eq.(\ref{eq:Heta}), the \textit{adiabatic protocol} for the QIP on the Keldysh contour can be established by making the following substitutions:
\begin{align}
\Hop_0 &\leftarrow \Himp(0) + \Hbath; \\
\Hop_1(t) &\leftarrow \Himp^{\eta}(t)+ \Hhyb,
\end{align}
where $\Himp^{\eta}(t) = \Himp(t) - \Himp(0)$ for $t\geq 0$ and $0$ otherwise.
However, the adiabatic protocol is not efficient for numerical calculation: the bare impurity dynamics determined by $\Himp^{\eta}(t)$ is time-dependent throughout the whole time interval $[-\infty, +\infty]$, which completely breaks the time-translational invariance. As a result, to evaluate any multi-time correlations of the impurity (which are the primary observables of interest for QIPs), one has to traverse the whole real-time axis, even though the Feynman-Vernon IF is time-translationally invariant. 

To avoid such complexity, we make the following \textit{equilibration assumption}: \textit{if we couple an impurity to a thermal bath with inverse temperature $\beta$, then they will reach equilibrium with inverse temperature $\beta$ after infinitely long time} (of course the scale of the impurity should be negligible compared to the bath). Mathematically, this assumption means:
\begin{align}\label{eq:assumption}
e^{-\beta\Hop(0)} \propto \lim_{t\rightarrow \infty} e^{-\im \Hop(0) t} \rhoop_{\rm imp}\otimes e^{-\beta\Hbath} e^{\im \Hop(0) t},
\end{align}
where the choice of the impurity initial state $\rhoop_{\rm imp}$ is irrelevant.
We note that the equilibration assumption, although sounds quite intuitive, has not been rigorously proven to our knowledge. In fact, it is much stronger than the adiabatic assumption stated in Sec.~\ref{sec:PI}: it is equivalent to stating that for QIPs, 
Eq.(\ref{eq:adiabatic}) holds even if $\eta(t)$ is the Heaviside step function (e.g., $\eta(t)=0$ for $t<0$ and $1$ otherwise). 
In the extreme case where the bath contains a single fermionic mode, the equilibration assumption will certainly fail, but the adiabatic assumption could still hold.
\gcc{We believe that this assumption is valid as long as we are dealing with a continuous bath whose energy scale is much larger than the impurity. Till now this assumption remains valid in all our numerical calculations. But we are not able to prove it rigorously. So it is left as an open question for future theoretical investigations.}
In addition, assuming that we are interested in calculating the multi-time impurity correlations within the time window $[0, t_f]$, then after time $t_f$ we perform another quench from $\Himp(t_f)$ to $\Himp(0)$. Due to the causality property of real time dynamics, this second quench will have no observable effects before time $t_f$, but it will be very useful for our numerical implementation. 

In summary, to solve the non-equilibrium QIP with time-dependent impurity Hamiltonian on the equivalent Keldysh contour, we will choose the following protocol for the time-dependent Hamiltonian:
\begin{align}\label{eq:Himp}
\Hop^{\eta}(t) = \begin{cases}
\Himp(t) + \Hint, &\text{if } t \in [0, t_f]; \\
\Himp(0) + \Hint, &\text{otherwise},
\end{cases}
\end{align}
with the initial state at $t=-\infty$ to be a separable state.
This protocol will be equivalent to the original problem on the Kadanoff-Baym contour as long as the equilibration assumption is valid.




\section{The non-equilibrium iGTEMPO method}
In the following we restrict our discussions to the single-orbital Anderson impurity model with
\begin{align}
\Himp(t) &= \varepsilon_d(t) \sum_{\sigma = \uparrow\downarrow} \adop_{\sigma}\aop_{\sigma} + U(t)\adop_{\uparrow}\adop_{\downarrow}\aop_{\downarrow}\aop_{\uparrow}; \\
\Hhyb &= \sum_{k, \sigma}V_k(\adop_{\sigma}\cop_{k, \sigma} + \hc); \\
\Hbath &= \sum_{k, \sigma} \varepsilon_{k}\cdop_{k, \sigma}\cop_{k, \sigma},
\end{align}
but the proposed method is applicable for any time-dependent AIMs as long as $\Hint$ remains time-independent. Here $\adop_{\sigma}, \aop_{\sigma}$ and $\cdop_{\sigma,k}, \cop_{\sigma,k}$ are the fermionic creation and annihilation operators of the impurity and bath respectively,  $\varepsilon_d(t)$ and $U(t)$ are the time-dependent on-site energy and interaction strength, $\varepsilon_{k}$ is the band energy, and $V_k$ is the coupling strength between the impurity and bath.

\subsection{The path integral formalism}
The impurity path integral of the time-dependent single-orbital AIM on the Keldysh contour, defined as
$\Zimp\EqDef\Tr\rhoop(\infty)/\Tr \rhoop_{\mathrm{bath}}^{\mathrm{th}}$, can be written in terms of Grassmann trajectories as~\cite{kamenev2009-keldysh,negele1998-quantum}
\begin{align}\label{eq:PI}
\Zimp = \int \mathcal{D}[\boldabar,\bolda] \gK\left[\boldabar, \bolda \right]\prod_{\sigma}\gI_{\sigma}\left[\boldabar_{\sigma}, \bolda_{\sigma}\right],
\end{align}
where  $\boldabar_{\sigma} = \{\abar_{\sigma}(\tau)\}$, $\bolda_{\sigma} = \{a_{\sigma}(\tau)\},\boldabar=\{\boldabar_{\uparrow},\boldabar_{\downarrow}\},\bolda=\{\bolda_{\uparrow},\bolda_{\downarrow}\}$ for briefness. The measure $\mathcal{D}[\boldabar,\bolda]=\prod_{\sigma,\tau}\dd\abar_{\sigma}(\tau)\dd a_{\sigma}(\tau)
  e^{-\abar_{\sigma}(\tau)a_{\sigma}(\tau)}$ ($\abar$ and $a$ are conjugate Grassmann variable of each other).
$\gK[\boldabar, \bolda]$ denotes the contribution from the bare impurity Hamiltonian which can be written as
\begin{align}
\gK[\boldabar, \bolda] = e^{-\im \int_{\contour}\dd\tau\gHimp(\tau)}.
\end{align}
Here $\contour = \contour^+ \cup \contour^-$ denotes the Keldysh contour, $\gHimp(\tau)$ is obtained from $\Himp(\tau)$ by making the substitutions $\aop_{\sigma}(\tau)\rightarrow a_{\sigma}(\tau)$ and $\adop_{\sigma}(\tau)\rightarrow\abar_{\sigma}(\tau)$, $\dd\tau$ should be understood as branch-dependent, e.g., $\dd\tau=\pm\delta t$ on $\contour^{\pm}$ with $\delta t$ the time step size on the real-time axis.
$\gI_{\sigma}[\boldabar_{\sigma}, \bolda_{\sigma}]$ denotes the Feynman-Vernon IF for a single spin species:
\begin{align}\label{eq:I}
\gI_{\sigma}[\boldabar_{\sigma}, \bolda_{\sigma}] = e^{-\int_{\contour} \dd\tau \int_{\contour}\dd\tau' \abar_{\sigma}(\tau) \Delta(\tau, \tau') a_{\sigma}(\tau') }.
\end{align}
The hybridization function $\Delta(t,t')$ encodes all the bath effects and can be calculated by
\begin{align}
  \label{eq:hybridization}
  \Delta(t,t')=\im\int\dd{\varepsilon}J(\varepsilon)D_{\varepsilon}(t,t'),
\end{align}
with $J(\varepsilon) = \sum_kV_k^2\delta(\varepsilon - \varepsilon_k)$ the bath spectrum density,
$D_{\varepsilon}(t,t')$ the free bath contour-ordered Green's function:
\begin{align}
  D_{\varepsilon}(t,t')\EqDef-\im\expval*{T_{\branch}\cop_{\varepsilon}(t)\cdop_{\varepsilon}(t')}_{\mathrm{bath}}.
\end{align}
Here $T_{\branch}$ is the contour-ordering operator that arranges operators on the contour in the order indicated by the arrows in Fig.~\ref{fig:fig1}(b), and $\expval{\cdots}_{\mathrm{bath}}$ means the expectation value with respect to the free bath.

The first step in GTEMPO is to discretize the impurity path integral using the quasi-adiabatic propagator path integral (QuAPI) method~\cite{makarov1994-path,makri1995-numerical}. Using a discrete time step size as $\delta t$, $\gI_{\sigma}$ can be written as (we will use the first-order discretization scheme throughout this work)~\cite{GuoChen2024e}
\begin{align}
\gI_{\sigma} = e^{-\sum_{\zeta, \zeta' = \pm}\sum_{j, k=-\infty}^{\infty} \abar_{\sigma, j}^{\zeta}\Delta_{j, k}^{\zeta\zeta'}a_{\sigma, k}^{\zeta'}},
\end{align}
where $\zeta, \zeta'$ are branch labels, $\Delta_{j, k}^{\zeta \zeta'}$ is the discrete hybridization function. As $\Delta_{j, k}^{\zeta \zeta'}$ is a function of the time difference $j-k$ only (which is essentially because that we consider time-independent $\Hint$), $\gI_{\sigma}$ is time-translationally invariant, which closely resembles the thermal state of an 1D infinite quantum manybody system~\cite{GuoChen2024d}. 
Therefore one could follow exactly the same procedure in Ref.~\cite{GuoChen2024e} to build $\gI_{\sigma}$ as an infinite GMPS, and these details will not be repeated in this work.

\subsection{The bare impurity dynamics}

\begin{figure}
  \includegraphics[width=\columnwidth]{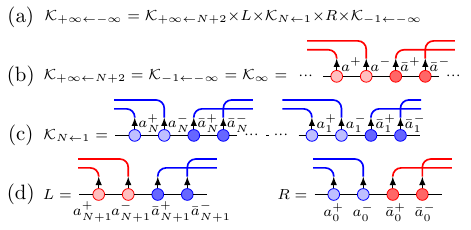} 
  \caption{Under the equilibration assumption in Eq.(\ref{eq:Himp}), the bare impurity dynamics can be represented as a window GMPS in (a), e.g., a finite GMPS $\gK_{N\leftarrow 1}$ in (c) bordered by two infinite GMPSs $\gK_{+\infty\leftarrow N+2}$ and $\gK_{-1\leftarrow -\infty}$ in (b). Both $\gK_{+\infty\leftarrow N+2}$ and $\gK_{-1\leftarrow -\infty}$ can be denoted as $\gK_{\infty}$ as they are the same. The two leads $L$ and $R$ in (d) are used to interface the finite GMPS with the two infinite GMPSs. The blue solid lines mean to apply two-body operations (e.g., multiplication by a Grassmann tensor of two GVs, determined by $\Himp$~\cite{ChenGuo2024a}) on $\gK_{N\leftarrow 1}$, while the red solid lines mean to apply two-body operations on $\gK_{\infty}$. Here we have removed the spin indices for notational simplicity.
    }
    \label{fig:fig2}
\end{figure}

\begin{figure*}
  \includegraphics[width=2\columnwidth]{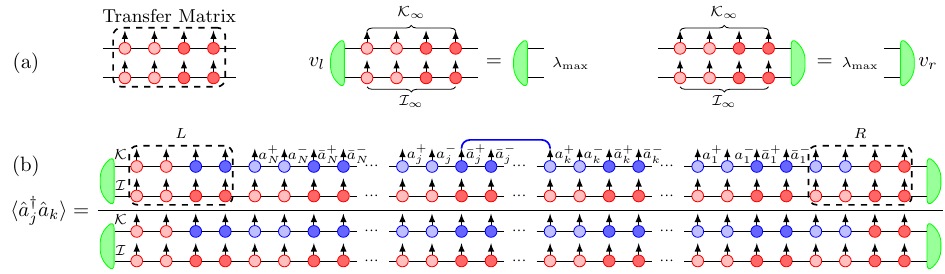} 
  \caption{The zipup algorithm to calculate Green's functions based on the infinite GMPS representation of $\gI$ (denoted as $\gI_{\infty}$) and the window GMPS representation of $\gK_{+\infty\leftarrow -\infty}$, which can be implemented with two steps: (a) calculating the left and right dominant eigenstates of the transfer matrix formed by integrating the product of $\gK_{\infty}$ and $\gI_{\infty}$ (which is performed on the fly) within a single time step; (b) evaluating the Green's function similar to the finite case, but using $L\times \gK_{N\leftarrow 1}\times R$ as the finite GMPS for the bare impurity dynamics, and with nontrivial left and right boundary tensors $v_l$ and $v_r$ obtained from step (a). 
  The blue solid line on the numerator of (b) means a two-body operation on $\gK_{N\leftarrow 1}$.
  The only difference with the zipup algorithm used for time-independent Hamiltonian~\cite{GuoChen2024e} is that in this case we have used a finite GMPS in the window of panel (b), which is different from that used for calculating the boundary tensors in panel (a). Again the spin indices are removed for notational simplicity.
    }
    \label{fig:fig3}
\end{figure*}

The efficient treatment of $\gK$ is more involved in the non-equilibrium case. 
In Ref.~\cite{GuoChen2024e}, one considers a time-independent total Hamiltonian, therefore both $\gK$ and $\gI$ are represented as infinite GMPSs when targeting at the steady state. In our case, if we use the adiabatic protocol
in Eq.(\ref{eq:adiabatic}), then there is no time-translational invariance in $\gK$, and one has to treat $\gK$ as a finite GMPS over the whole time interval $[-\infty, \infty]$ (again this could be slightly shortened to $[-\infty, t_f]$ due to the causality property). Moreover, to calculate any impurity correlation, one has to traverse the whole real-time axis which would wipe away any computational advantage of using an infinite $\gI_{\sigma}$.

Fortunately, if one uses the equilibration assumption in Eq.(\ref{eq:Himp}), then $\gK$ can be built as a \textit{window GMPS}, which is a finite GMPS bordered by two infinite GMPSs on the two sides. The structure of the whole $\gK_{+\infty \leftarrow -\infty}$ is schematically illustrated in Fig.~\ref{fig:fig2}, where we have neglected the spin indices. The two infinite GMPSs represent the bare impurity dynamics formed by evolving the impurity with $\Himp(0)$ for infinitely long time, while the finite GMPS $\gK_{N\leftarrow 1}$ represents the impurity dynamics formed by evolving the impurity with $\Himp(t)$ from $0$ to $t_f$ ($N = t_f/\delta t$). In addition, we have used two leads, denoted by $L$ and $R$, to interface the finite GMPS with the two infinite GMPSs on the two sides, which is crucial for numerical implementation as otherwise the auxiliary states of the finite GMPS and the infinite GMPSs will mismatch. 
In Fig.~\ref{fig:fig2} we have ordered the discrete Grassmann variables (GVs) within each time step as $a_j^+ a_j^- \abar_j^+ \abar_j^-$, which is the most convenient ordering for the GMPS representation of $\gK$. In our actual implementation, however, we still stick to the ordering of GVs used in Ref.~\cite{GuoChen2024e} where the conjugate pairs of GVs are put in neighbouring positions (the latter ordering is the most convenient for integrating out the GVs in the end to calculate observables). The two different orderings can be easily transformed into each other using local swap gates.

Importantly, the only computational overhead, compared to the case of time-independent Hamiltonian, is that one needs to build a finite GMPS representation $\gK_{N\leftarrow 1}$ on top of the infinite $\gK_{\infty}$, the cost of which is negligible since the bond dimension of $\gK_{N\leftarrow 1}$ is a small fixed number as long as the \textit{time-local ordering} (GVs within the same time step are located nearby) of GVs is used. 

Overall, the computational cost to build the GMPS representations of the impurity path integral for time-dependent AIMs is roughly the same as that for time-independent AIMs, as the dominate computational cost is to build the Feynman-Vernon IF as an infinite GMPS which is the same in both cases. For building the infinite GMPS representation of the Feynman-Vernon IF, the dominate calculation is the multiplication of two infinite GMPSs (which is analogous to the element-wise product of two normal arrays) followed by the canonicalization of the result infinite GMPS. For these operations there exists well-established infinite MPS techniques~\cite{OrusVidal2008,StauberHaegeman2018}, and their computational costs are essentially independent of the real or imaginary time.

\subsection{Calculating Green's functions }

Once $I_{\sigma}$ are built as infinite GMPSs and $\gK_{+\infty \leftarrow -\infty}$ is built as a window GMPS, one could calculate Green's functions (or generally any multi-time impurity correlations) within the time window $[0, t_f]$ using the same zipup algorithm in Refs.~\cite{GuoChen2024e,GuoChen2024f}, which is illustrated in Fig.~\ref{fig:fig3}.
Compared to the time-independent case, the only difference is: for time-dependent $\Himp(t)$ we have used a finite GMPS representation of $\Himp(t)$ (plus two leads) within the time window and used $\Himp(0)$ for calculating the leading eigenstates on the two sides, while in the time-independent case we have used $\Himp$ both within the time window and for calculating the leading eigenstates. Given $I_{\sigma}$ and $\gK_{+\infty \leftarrow -\infty}$, the computational cost for calculating the Green's functions scales as $O(N)$ (e.g., linearly against the window size $t_f$) as can be seen from Fig.~\ref{fig:fig3}(b). Importantly, by properly caching the intermediate calculations (similar to the caching of environments in general MPS algorithms~\cite{Schollwock2011}), the cost of calculating all the Green's functions within the time window only scales linearly as $t_f$. In addition, the prefactor of this linear scaling is so low that we can easily reach a very large $t_f$ when calculating the Green's functions. We will give some concrete runtimes in our calculations later.

\section{Numerical results}
In this section we demonstrate the performance of the proposed non-equilibrium infinite GTEMPO method in both the noninteracting and interacting cases.
\gcc{For both cases, we will consider two different inverse temperatures: $\Gamma\beta=4$ and $\beta=\infty$ (zero temperature) to illustrate the flexibility of our method.}
For all the numerical calculations in this work, we adopt the following bath spectral density
\begin{equation}
    J(\omega) = \frac{\Gamma D}{2\pi} \sqrt{1-(\omega/D)^2}
\end{equation}
with $D=2$ and $\Gamma=0.1$ (the same bath spectral density has also considered in Refs.~\cite{ThoennissAbanin2023b,ChenGuo2024a,BertrandWaintal2019}). 
We will take $\Gamma$ as the unit. 
Although our method is equally applicable for any time-dependent impurity Hamiltonian, in our numerical simulations we will focus on the case that the impurity Hamiltonian undergoes a sudden quench at $t=0$.

\subsection{The noninteracting case}

\begin{figure}[!h]\centering
  \includegraphics[width=\columnwidth]{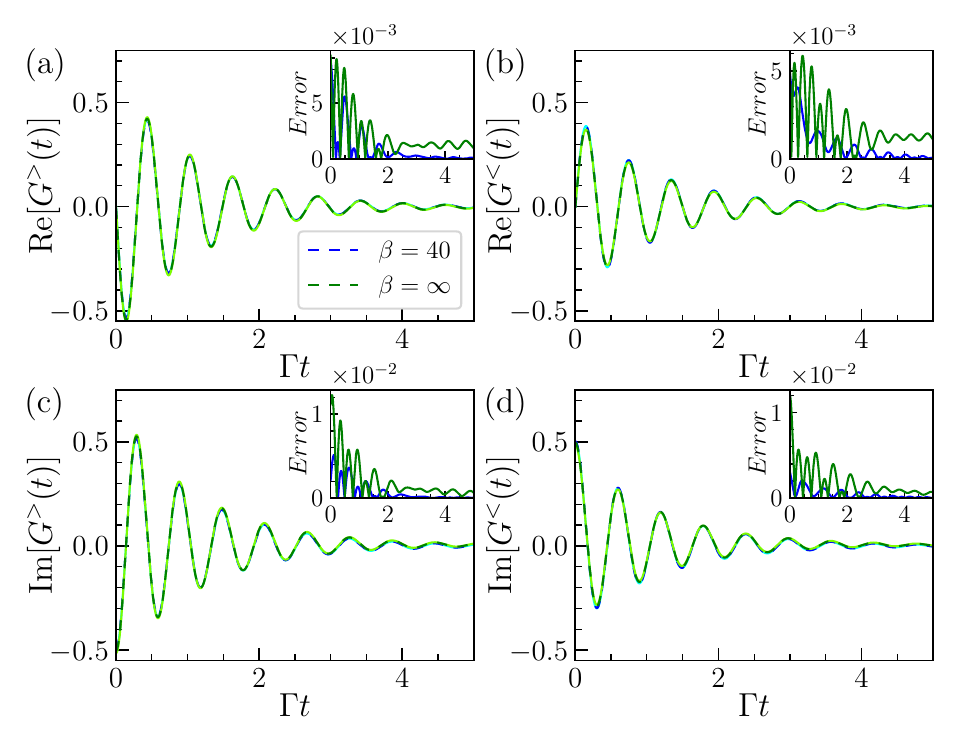}
  \caption{(a) The real and (c) imaginary parts of the non-equilibrium greater Green's function $G^>(t)$ for the Toulouse model. (b) The real and (d) imaginary parts of the non-equilibrium lesser Green's function $G^<(t)$. The blue \gcc{and green} dashed lines are neq-iGTEMPO results \gcc{for $\Gamma\beta=4$ and $\beta=\infty$ respectively,} while the solid lines \gcc{with the same colors} are \gcc{the corrsponding} ED results. The insets show the errors between the neq-iGTEMPO results and the ED results. \gcc{For the neq-iGTEMPO results, We have used $\chi=48$ at $\Gamma\beta=4$, and $\chi=60$ at $\beta=\infty$, $\Gamma\delta t=0.005$ is used for both temperatures.} 
  }
    \label{fig:toulouse1}
\end{figure}

\begin{figure}[!h]\centering
  \includegraphics[width=\columnwidth]{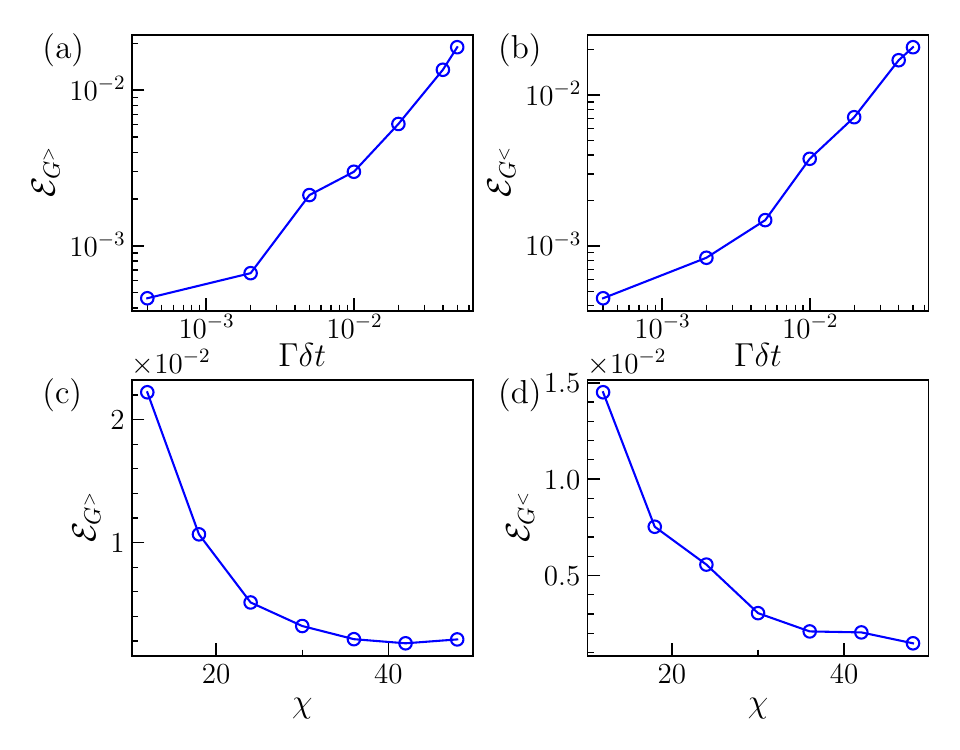}
  \caption{(a, b) The average errors of (a) the greater and (b) the lesser Green's functions calculated by neq-iGTEMPO against ED as functions of $\delta t$ for the Toulouse model, where we have fixed $\chi=48$. (c, d) The average errors of (c) the greater and (d) the lesser Green's functions calculated by neq-iGTEMPO against ED as functions of $\chi$, where we have fixed $\Gamma\delta t=0.005$ \gcc{and $\Gamma\beta=4$}. }
    \label{fig:toulouse2}
\end{figure}

\begin{figure}[!h]\centering
  \includegraphics[width=\columnwidth]{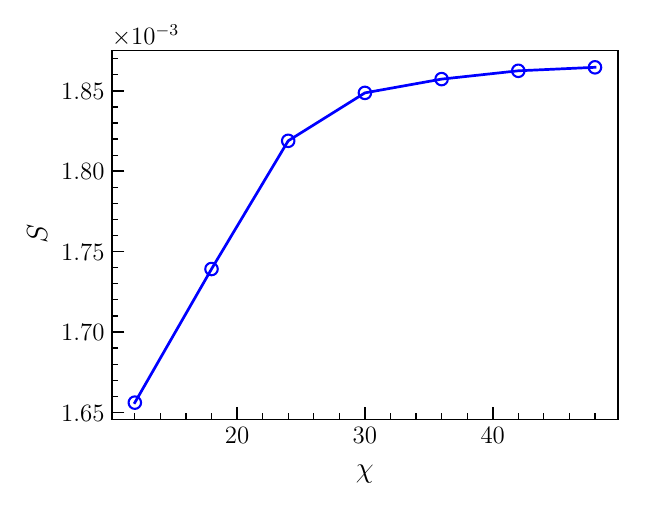}
  \caption{The bipartition entanglement entropy $S$ of the Feynman-Vernon IF $\gI_{\sigma}$ as a function of $\chi$. Here we have used $\Gamma\delta t=0.005$ \gcc{and $\Gamma\beta=4$}.}
    \label{fig:toulouse3}
\end{figure}

We first consider the noninteracting case, which is integrable and referred to as the Toulouse model. We consider the following non-equilibrium scenario: the impurity is initially in thermal equilibrium with the bath under $\varepsilon_d=0$ (half filling), and then the on-site energy is suddenly quenched to $\varepsilon_d/\Gamma=10$ for $t\geq 0$.


In Fig.~\ref{fig:toulouse1}, we compare the greater and lesser Green's functions obtained from neq-iGTEMPO to those from ED, \gcc{which are defined as:
\begin{align}
G^>_{j} &= G^>(j\delta t) = -\im \langle \aop_{\sigma, j}\adop_{\sigma, 0} \rangle; \\
G^<_{j} &= G^<(j\delta t) = \im \langle \adop_{\sigma, 0}\aop_{\sigma, j}\rangle.
\end{align}
The blue dashed and solid lines repesent the neq-iGTEMPO and ED results for $\Gamma\beta=4$ respectively, while the green dashed and solid lines are for  $\beta=\infty$.}
For the neq-iGTEMPO calculations, We have chosen a discrete time step size of $\Gamma\delta t=0.005$ and a maximally allowed bond dimension $\chi=48$ for $\gI_{\sigma}$. The ED results are computed with a bath discretization of $\delta\omega/\Gamma=0.1$ (which is the only source of error in ED), and have well converged in all our tests. 
We can see that the neq-iGTEMPO results agree well with ED for both $G^>$ and $G^<$ \gcc{and for both temperatures (these two sets of results are very close to each other, which is a coincidence in thise case)}. 
The insets show the errors between the neq-iGTEMPO results and the ED results, which are on the order of $10^{-3}$ and do not grow with time \gcc{(the errors for $\beta=\infty$ is only slightly larger than those for $\Gamma\beta=4$)}. This error behavior is expected since in PI-based MPS methods the whole time interval is treated on the same footing.

In Fig.~\ref{fig:toulouse2}, we check the convergence of the neq-iGTEMPO calculations against the two important hyperparameters, $\delta t$ and $\chi$, for the Toulouse model \gcc{with $\Gamma\beta=4$} (There is an additional source of error in the iGTEMPO algorithm compared to the GTEMPO algorithm, that is, the error occurred in the Prony algorithm to approximate the hybridization function as the sum of exponentially decaying functions. However, this error only happens in the preprocessing of the hybridization function and can be made very small in principle, without significant effect on cost of the iGTEMPO calculations. In all our numerical simulations we have chosen a large enough expansion order for the Prony algorithm such that the relative error is smaller than $10^{-3}$. One can also see Ref.~\cite{GuoChen2024d} for details of the Prony algorithm). 
We use the average error, defined as
\begin{align}
    \mathcal{E}(\vec{x},\vec{y}) = \sqrt{\frac 1N \sum_{i=1}^N{{|x_i-y_i|}^2}}
\end{align}
between two vectors $\vec{x}$ and $\vec{y}$, to access the overall error of the neq-iGTEMPO results compared to the ED results.
From Fig.~\ref{fig:toulouse2}(a,b), we fix $\chi=48$ and study the average error as a function of $\delta t$. We observe that $\mathcal{E}$ decreases with smaller time step $\delta t$. In particular, $\mathcal{E}$ decreases approximately linearly when $\Gamma\delta t > 0.01$, indicating that the time discretization error is the dominant source of error in this regime. 
In Fig.~\ref{fig:toulouse2}(c,d), we fix the time step size to $\Gamma\delta t=0.005$ and study the average error as a function of $\chi$. We observe that $\mathcal{E}$ decreases and converges to a non-zero value (which is due to the finite time discretization error). 
Interestingly, with a small bond dimension $\chi=30$, we can already obtain fairly accurate results, which demonstrates the efficiency of the neq-iGTEMPO method.

The saturation of the error with respect to the bond dimension $\chi$ means that the Feynman-Vernon IF $\gI_{\sigma}$ possesses only a finite amount of entanglement. Concretely, in Fig.~\ref{fig:toulouse3}, we plot the largest bipartition entanglement entropy $S$ of all bonds within a unit cell of $\gI_{\sigma}$ as a function of $\chi$, where we can also clearly see that $S$ quickly saturates with $\chi$.

To this end, we note that the computational costs of GTEMPO or iGTEMPO are directly related to the augmented density tensor (ADT), defined as $\gA[\boldabar, \bolda]=\gK[\boldabar, \bolda]\sum_{\sigma}\gI_{\sigma}[\boldabar, \bolda]$, as in principle any impurity observables can only be calculated based on $\gA$. 
As GMPS multiplication is similar to the tensor product of MPSs~\cite{GuoChen2024d}, in the worst case, the bipartition entanglement entropy of $\gA$ will be the summation of those of $\gK$ and $\gI_{\sigma}$ (therefore the computational cost will grow exponentially with the size of the impurity).
However, in our zipup algorithm (Fig.~\ref{fig:fig3}), the ADT is never explicitly built, but only calculated on the fly where we have also assumed this worst case behavior. As a result Fig.~\ref{fig:toulouse3} only reflects the worst case behavior of the bipartition entanglement entropy of the ADT. In future investigations it would be interesting to explore the possibility of explicitly building the ADT and compressing it without significant loss of accuracy.



\subsection{The single-orbital Anderson impurity model}

\begin{figure*}[htbp]
  \includegraphics[width=2\columnwidth]{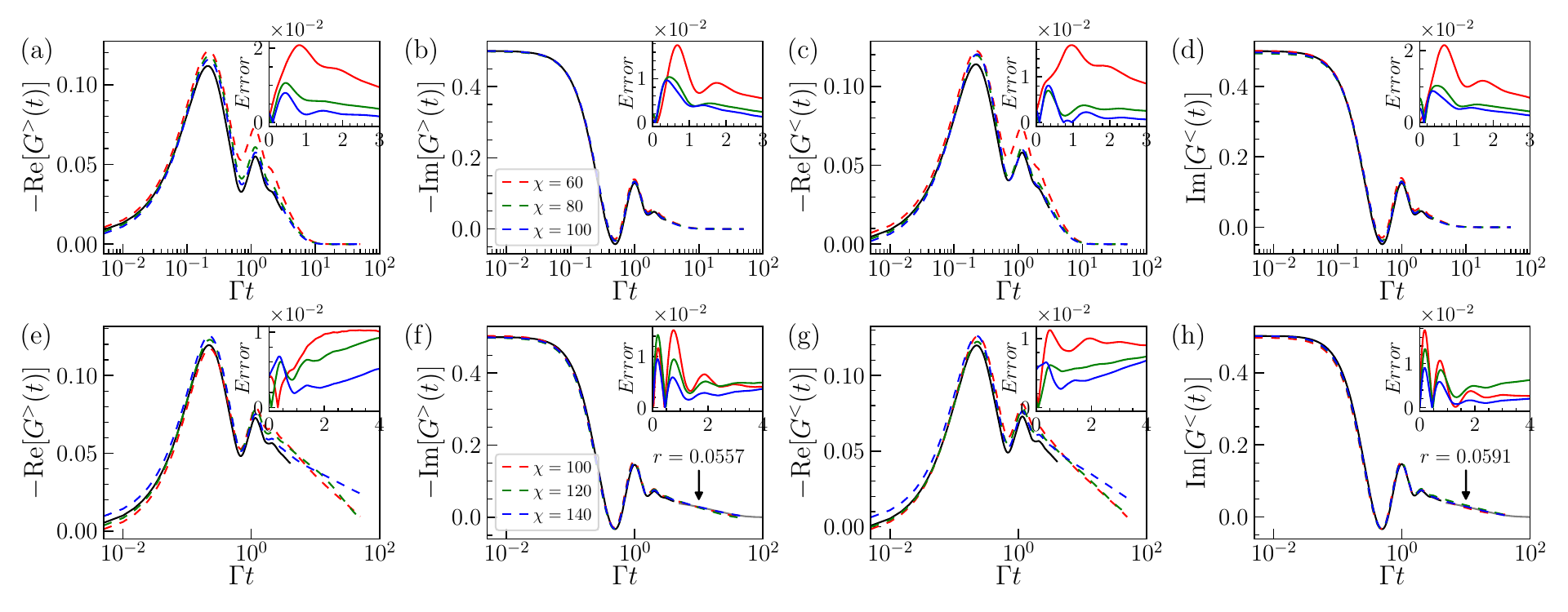} 
  \caption{\gcc{(a) The real and (b) the imaginary parts of $G^>$ for the single-orbital AIM with $\Gamma\beta=4$. (c) The real and (d) the imaginary parts of $G^<$ for the single-orbital AIM with $\Gamma\beta=4$. The black solid lines represent the GTEMPO results performed on the L-shaped Kadanoff-Baym contour using $\chi=160$. The red, green, and blue dashed lines correspond to the neq-iGTEMPO results with $\chi=60,80,100$ respectively. (e,f,g,h) Similar to (a,b,c,d), but for $\beta=\infty$. The red, green, and blue dashed lines correspond to the neq-iGTEMPO results with $\chi=100,120,140$ respectively. The black solid lines represent the GTEMPO results performed on the Keldysh contour using $\Gamma t_0=6$ and $\chi=180$. 
  The gray solid lines in (f,h) show the exponential fitting of the tail after $\Gamma t=5$ with decay rate $r$.
  The insets in all the panels show the errors between the neq-iGTEMPO results and the GTEMPO results, where the red, green and blue solid lines correspond to the neq-iGTEMPO results in the main panel with the same colors respectively.}}
    \label{fig:SISB1}
\end{figure*}

\begin{figure*}[!h]\centering
  \includegraphics[width=2\columnwidth]{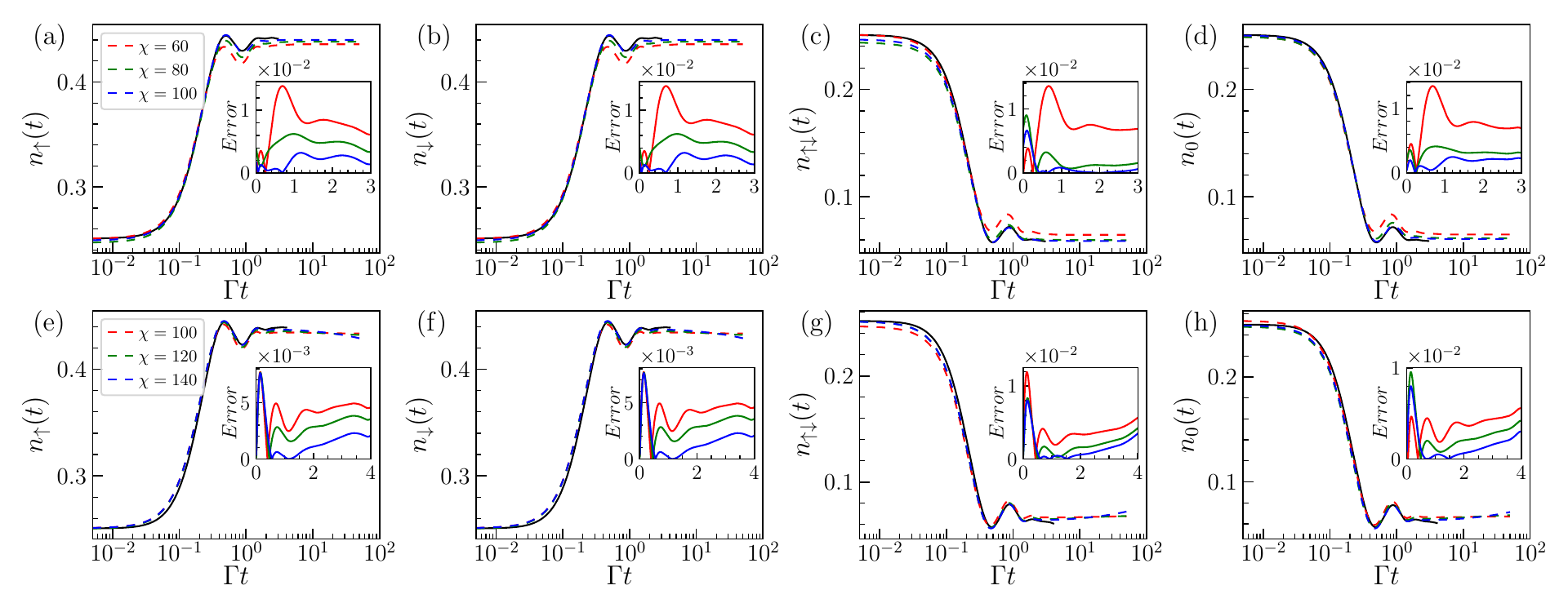}
  \caption{\gcc{(a,b,c,d) The impurity populations in the four states: (a) $\ket{\uparrow}$, (b) $\ket{\downarrow}$, (c) $\ket{\uparrow\downarrow}$, (d) $\ket{0}$, as functions of time for the single-orbital AIM \gcc{with $\Gamma\beta=4$}. The solid lines represent the GTEMPO results performed on the L-shaped Kadanoff-Baym contour using $\chi=160$. The red, green, and blue dashed lines correspond to the neq-iGTEMPO results with $\chi=60,80,100$ respectively. 
  (e,f,g,h) Similar to (a,b,c,d), but for $\beta=\infty$.  The solid lines represent the GTEMPO results performed on the Keldysh contour using $\Gamma t_0=6$ and $\chi=180$. The red, green, and blue dashed lines correspond to the neq-iGTEMPO results with $\chi=100,120,140$ respectively. 
  The insets show the errors between the neq-iGTEMPO results and the corresponding GTEMPO results. }
  }
    \label{fig:SISB2}
\end{figure*}

\begin{figure}[!h]\centering
  \includegraphics[width=\columnwidth]{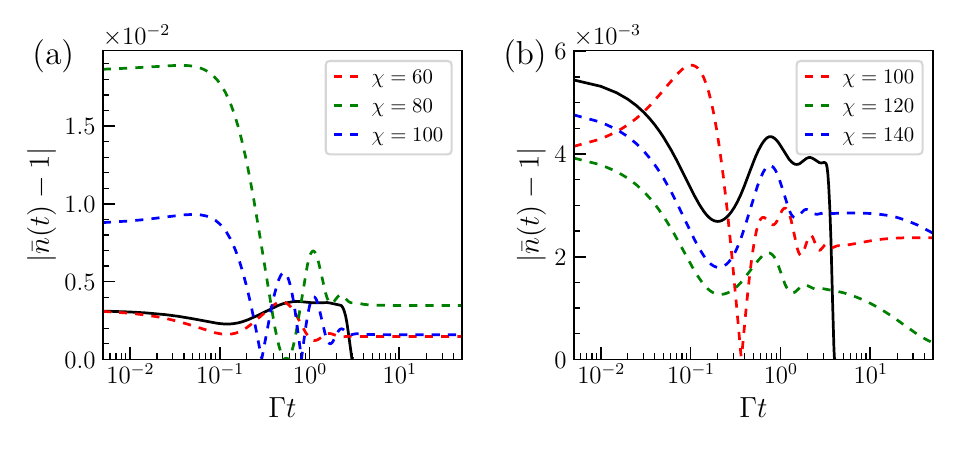}
  \caption{Deviation of the average occupation $\bar n$ from half filling for (a) $\Gamma\beta=4$ and (b) $\beta=\infty$. \gcc{In panel (a), the black solid line is the GTEMPO results performed on the L-shaped Kadanoff-Baym contour using $\chi=160$, the red, green and blue dashed lines correspond to the neq-iGTEMPO results calculated with $\chi=60, 80, 100$ respectively. In panel (b), the black solid line is the GTEMPO results performed on the Keldysh contour using $\Gamma t_0=6$ and $\chi=180$, the red, green and blue dashed lines correspond to the neq-iGTEMPO results calculated with $\chi=100, 120, 140$ respectively.}
  }
    \label{fig:SISB3}
\end{figure}

In the next we consider the single-orbital AIM.
We focus on the half-filling case with $\varepsilon_d=-U/2$.
We consider the following non-equilibrium scenario: the impurity is initially in equilibrium with the bath under $U=0$, and then we suddenly turn on $U/\Gamma = 10$ for $t\geq 0$. \gcc{Here We note that the choice $U=0$ made for the initial state is not a simplification for our method, we can choose the initial equilibrium state under any parameter settings of the impurity Hamiltonian with the same effort.}
As this model is not integrable and can not be solved using ED, we will benchmark the neq-iGTEMPO results against GTEMPO calculations. For both methods we fix $\Gamma\delta t=0.005$. For GTEMPO calculations we will use a large bond dimension, such that the GTEMPO results suffer less from the MPS bond truncation error and can be used as a proper benchmarking baseline (the GTEMPO results still suffer from the first-order time discretization error in QuAPI).

In Fig.~\ref{fig:SISB1}(a,b,c,d), we show the real and imaginary parts of $G^>$ and $G^<$ \gcc{for $\Gamma\beta=4$}, where the black solid lines represent the GTEMPO results \gcc{performed on the L-shaped Kadanoff-Baym contour~\cite{ChenGuo2024g}} with $\chi=160$, while the red, green, blue dashed lines correspond to the neq-iGTEMPO results with $\chi=60,80,100$ respectively. We have used a large time window $\Gamma t_f = 50$ for neq-iGTEMPO, and used $\Gamma t_f = 3$ for GTEMPO due to the limitation of computational efficiency in the latter case. We have also used log scale for the x-axis to help visualizing the long time behavior.
We can see that the neq-iGTEMPO results generally agree well with the GTEMPO results for $\Gamma t \leq 3$. For $\Gamma t > 3$, the neq-iGTEMPO results for different $\chi$s are still well on top of each other.
\gcc{In Fig.~\ref{fig:SISB1}(e,f,g,h), we plot the same observables as those plotted in Fig.~\ref{fig:SISB1}(a,b,c,d), but for $\beta=\infty$ instead, and the red, green, blue dashed lines correspond to the neq-iGTEMPO results with $\chi=100,120,140$ respectively. In this case, it is not possible to perform GTEMPO calculations on the L-shaped contour as $\beta$ is infinite. To benchmark our neq-iGTEMPO results in this case, we take the idea used in Ref.~\cite{ChenGuo2024c} which only requires GTEMPO calculations on the real-time axis (e.g., the Keldysh contour): we first evolve the whole system from a separable impurity-bath initial state for a long enough time $t_0$, where the bath is initially in thermal state and the impurity initial state is chosen as $\vert 0\rangle$, such that the impurity already equilibrates with the bath ($t_0$ is thus referred to as the \textit{equilibration time}), then we quench the impurity Hamiltonian and calculate the Green's functions (as well as other observables considered in the following) from $t_0$. As benchmark baseline, we have used GTEMPO results calculated with $\Gamma t_0=6$ and $\chi=180$, which are shown in black solid lines (see Appendix.~\ref{app:equilibration} for a convergence study of the GTEMPO results used here).
}
\gcc{In the insets, we show the errors between the neq-iGTEMPO results and baseline}, which are all on the order of $10^{-2}$ (about the same order as $\delta t$), demonstrating the consistency between these two methods that are based on very different but equivalent mathematical formulations. \gcc{For $\Gamma\beta=4$, the error decreases significantly as $\chi$ increases from $60$ to $80$, 
and then approximately saturate when $\chi$ is further increased to $100$. While for $\beta=\infty$ the signal of error saturation against the baseline is not as clear. In addition, for $\beta=\infty$, larger bond dimensions have been used, yet from Fig.~\ref{fig:SISB1}(e,g) we observe that the convergence is still not as good as in Fig.~\ref{fig:SISB1}(a,c), which indicates that the zero-temperature case is more challenging for our method)
In Fig.~\ref{fig:SISB1}(f,h), we further show the exponential fitting $e^{-r\Gamma t}$ (the gray solid lines) for the imaginary parts of the two Green's functions after $\Gamma t=5$, where the decay rate $r$ is around $0.06$ in both cases, close to the Kondo energy scale $0.062$~\cite{Georges2016} (see Appendix.~\ref{app:kondo} for the estimation of the Kondo energy scale).
We note that the fittings for the real parts of the two Green's functions are not shown, as they will cancel each other when calculating the retarded Green's function, defined as $G(t) = G^>(t)-G^<(t)$, under the half filling condition. Our fitting essentially aims for $G(t)$, which determines the spectral function.}


In Fig.~\ref{fig:SISB2}, we show the time evolution of the impurity populations in the four states $\ket{\uparrow}$ (spin up), $\ket{\downarrow}$ (spin down), $\ket{\uparrow\downarrow}$ (double occupancy), $\ket{0}$ (zero occupancy), and compare them to the GTEMPO calculations. \gcc{The top row shows the results for $\Gamma\beta=4$ and the bottom row for $\beta=\infty$.} We can see that the neq-iGTEMPO results agree very well with the GTEMPO results. The insets show the errors between the neq-iGTEMPO results and the GTEMPO results. \gcc{For these observables we observe that the error against the baseline decreases almost monotonically with larger $\chi$ for both temperatures.} 

As we have enforced the half filling condition throughout the time evolution (although $U$ has been changed), we expect the average electron occupation $\bar n = n_\uparrow + n_\downarrow + 2n_{\uparrow\downarrow}$ to be $1$ during the evolution, which can thus be used as a consistency check for our neq-iGTEMPO and GTEMPO calculations. In Fig.~\ref{fig:SISB3} we show $|\bar{n}-1|$ as a function of time \gcc{for both $\Gamma\beta=4$ (panel a) and $\beta=\infty$ (panel b)}, we can see that the deviation is on the order of $10^{-2}$ \gcc{or less} for all these results, therefore the half filling condition is indeed well preserved in our numerical calculations.

To this end, we give some concrete computation times in our numerical simulation of the single-orbital AIM \gcc{by taking the case $\Gamma\beta=4$ as an example} (we have used a single thread for all our calculations): Building the Feynman-Vernon IF $\gI_{\sigma}$ as an infinite GMPS takes around $50$ hours (currently we use the infinite density matrix renormalization group algorithm to compress the infinite GMPS, which is stable and accurate, but converges very slowly as it is essentially a power iteration method~\cite{GuoChen2024e}); Calculating the left and right dominate eigenpairs in Fig.~\ref{fig:fig3}(a) take around $1.6$ hours; Calculating the greater and lesser Green's functions in Fig.~\ref{fig:fig3}(b) with $\Gamma\delta t=0.005$ and $\Gamma t_f=50$ ($10000$ points each) takes around $2.2$ hours.


\section{Conclusion}

In summary, we have proposed an infinite Grassmann time-evolving matrix product operator method for non-equilibrium quantum impurity problems. By formulating the time-dependent impurity problem on an equivalent Keldysh contour, we restore the time-translational invariance of the Feynman-Vernon influence functional of the problem. Based on a specially designed quench protocol for the bare impurity dynamics, we can finally make full use of the infinite MPS technique in our method. The computational cost of the proposed method to build the Grassmann MPS representations of the impurity path integral is thus independent of the real or imaginary time, similar to the infinite GTEMPO method that aims at the steady state of time-independent impurity problems, greatly improving over the GTEMPO method on the L-shaped Kadanoff-Baym contour which scales with both the real and imaginary times.
Our method is ideal for studying long-time non-equilibrium dynamics of quantum impurity problems, and can be potentially used as an efficient impurity solver in the non-equilibrium DMFT.

\begin{acknowledgments}
Z. L. is partially supported by NSFC (22393913), by the Strategic Priority Research Program of the Chinese Academy of Sciences (XDB0450101). 
R. C. is supported by National Natural Science Foundation of China under Grant No. 12104328. 
C. G. is supported by the Open Research Fund from State Key Laboratory of High Performance Computing of China (Grant No. 202201-00). 
\end{acknowledgments}

\appendix

\section{Convergence study of the GTEMPO results at zero temperature}\label{app:equilibration}

\begin{figure*}[htbp]
  \includegraphics[width=2\columnwidth]{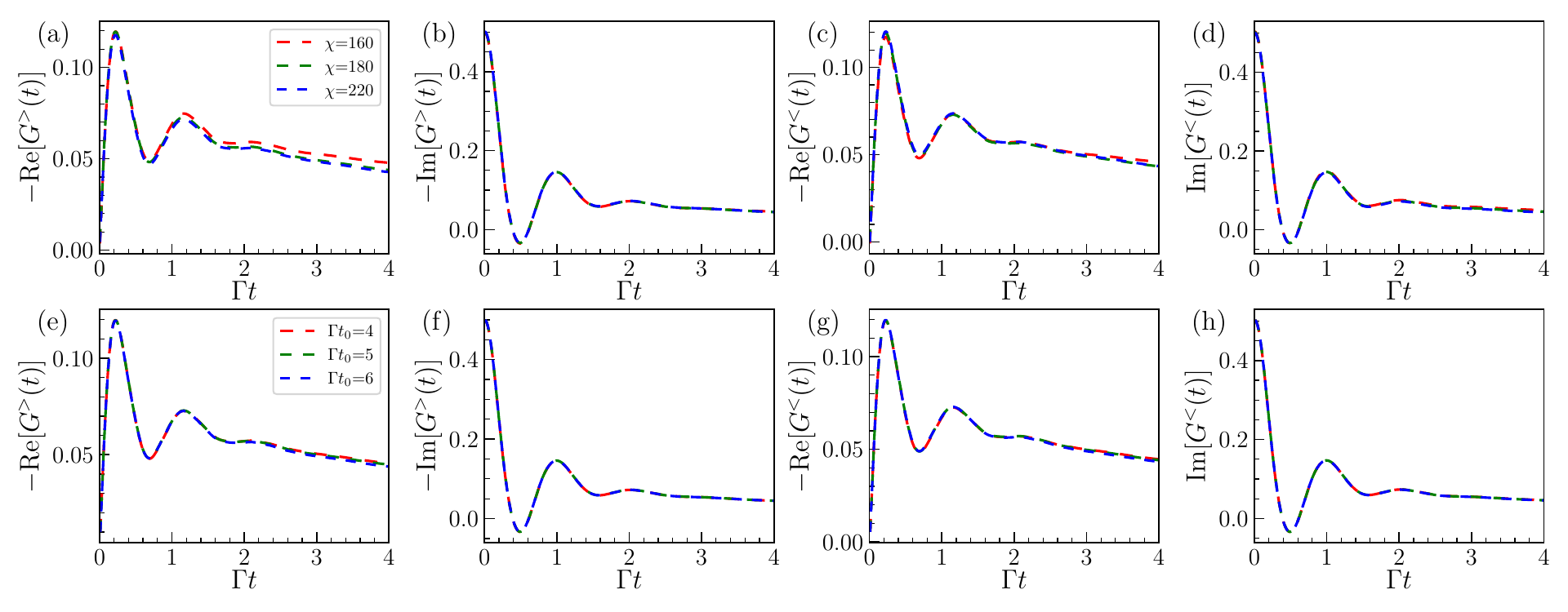} 
  \caption{Convergence study of GTEMPO calculations performed on the Keldysh contour for the single-orbital AIM considered in the main text, with $\beta=\infty$. In (a,b,c,d), we analyze the convergence against bond dimension $\chi$, where we have fixed $\Gamma t_0=6$ and the red, green, blue dashed lines are GTEMPO results calculated using $\chi=160,180,220$ respectively. In (e,f,g,h) we analyze the convergence against the equilibration time $t_0$, where we have fixed $\chi=180$, and the red, green, blue dashed lines are GTEMPO results calculated using $\Gamma t_0=4,5,6$ respectively. In these simulations we used $\Gamma\delta t=0.005$.
  }
    \label{fig:SISB_conv}
\end{figure*}

Here we show the convergence of our GTEMPO results performed on the Keldysh contour against two hyperparameters: the bond dimension $\chi$ and the equilibration time $t_0$, which is shown In Fig.~\ref{fig:SISB_conv}. For panels (a,b,c,d), we can see that the GTEMPO results calculated with $\chi=180$ well agree with those calculated with $\chi=220$, which indicates that the GTEMPO results have converged with $\chi=180$ under the given $\delta t$. From panels (a,b,c,d), we can see that the GTEMPO results calculated with $\Gamma t_0=4,5,6$ are all on top of each other, which indicates that with our choice of $\Gamma t_0=6$, the impurity have well reached equilibrium with the bath. 
These simulations thus qualify the GTEMPO calculations on the Keldysh contour using $\chi=180$ and $\Gamma t_0=6$ as a good benchmarking baseline.

\section{Estimating the Kondo energy scale at zero temperature}\label{app:kondo}
In Eq.(11) of Ref.~\cite{Georges2016}, the Kondo energy scale $T_K$ can be estimated as
\begin{align}
T_K = \sqrt{\pi UV^2\rho_0} e^{-\frac{1}{J_K\rho_0}},
\end{align}
where 
\begin{align}
J_K = 2V^2(\frac{1}{\varepsilon_d+U} - \frac{1}{\varepsilon_d} ),
\end{align}
and $V^2\rho_0 = J(0) = \Gamma/\pi$.
Under our parameter settings ($\varepsilon_d=-U/2$ and $U/\Gamma=10$), we have $J_K = \frac{4V^2}{5\Gamma}$, and therefore we have
\begin{align}
T_K = \sqrt{\pi U \times \frac{\Gamma}{\pi}}e^{-\frac{5\pi}{4}} \approx 0.062\Gamma.
\end{align}

\bibliographystyle{apsrev4-2}
\bibliography{refs}

\end{document}